\documentclass[aps,superscriptaddress,nofootinbib,eqsecnum,prd,notitlepage,twocolumn,showkeys]{revtex4-1} 

\pdfoutput=1

\usepackage{amsfonts}
\usepackage{amsmath}
\usepackage{amssymb}
\usepackage{graphicx,color}
\usepackage{float}
\usepackage{hyperref}
\usepackage{subfigure}
\usepackage{dcolumn}
\usepackage{soul}
\usepackage{ulem}
\usepackage{verbatim}
\usepackage[dvipsnames]{xcolor}
\def \violet {\color{violet}}

\begin{document}

\title{Witten-O'Raifeartaigh potential revisited in the context of Warm Inflation}

\author{Suratna Das}
\email{suratna.das@ashoka.edu.in}
\affiliation{Department of Physics, Ashoka University,
   Rajiv Gandhi Education City, Rai, Sonipat: 131029, Haryana, India}

\author{Umang Kumar}
\email{umang.kumar$_$phd21@ashoka.edu.in }
\affiliation{Department of Physics, Ashoka University,
   Rajiv Gandhi Education City, Rai, Sonipat: 131029, Haryana, India}

\author{Swagat S. Mishra}
\email{swagat.mishra@nottingham.ac.uk}
\affiliation{School of Physics and Astronomy, University of Nottingham, Nottingham, NG7 2RD, UK.}
\affiliation{Cosmology, Gravity, and Astroparticle Physics Group, Center for Theoretical Physics of the Universe (CTPU-CGA), Institute for Basic Science (IBS), Daejeon, 34126, Korea.}
   
 \author{Varun Sahni}
\email{varun@iucaa.in}
\affiliation{Inter-University Centre for Astronomy and Astrophysics, Post Bag 4, Ganeshkhind, Pune
411 007, India.}

\begin{abstract}

Warm Inflation is  a scenario in which the inflaton field dissipates its energy during inflation to maintain a subdominant constant radiation bath.  Two  of its remarkable features are (i) inflation can be realized even by very steep potentials and (ii) such a scenario doesn't call for a separate post-inflation reheating phase. We exploit the first feature to show that Warm Inflation can successfully take place on the very steep left wing of the Witten-O'Raifeartaigh potential while remaining in excellent agreement with current cosmological data (joint analysis of Planck, ACT and DESI). The  Witten-O'Raifeartaigh potential has a flatter right wing as well, which opens up the possibility of dark energy when the field rolls along this wing. However in order to successfully realize quintessential inflation one needs to (i) normalize the two wings of the Witten-O'Raifeartaigh potential differently in order to bridge between the two extreme energy scales of inflation and dark energy, (ii) allow the quintessence field to be dissipative, which is consistent with the presence of a dissipative term in warm inflation. The dissipative dynamics of the quintessence field is needed in order to sustain slow-roll in the right wing. With these modifications, we demonstrate that the Witten-O'Raifeartaigh potential can give rise to a unified model of warm inflation (on the left wing) and transient dark energy (on the right wing). 
\end{abstract}

\maketitle

\section{Introduction}

Cosmic inflation, an early accelerated phase of our universe, was originally introduced~\cite{Starobinsky:1980te, Kazanas:1980tx, Guth:1980zm, Albrecht:1982wi, Linde:1983psb} as an extension of the hot Big Bang model to overcome some of its severe finetuning shortcomings. This early accelerated epoch, though  motivated theoretically, has unique observational predictions (nearly-Gaussian, nearly-scale-invariant and nearly-adiabatic scalar primordial perturbations, and lesser tensor fluctuations compared to the scalar ones \cite{Baumann:2009ds}) that have been tested and confirmed by the current CMB observations \cite{Planck:2018jri}. The simple scenario of a single scalar field, a.k.a. the inflaton, slowly rolling down its moderately flat potential, can explain such an accelerated expansion of our universe. However, the rapid exponential expansion of  space during inflation quickly dilutes away all the other energy densities resulting in a supercooled universe  at the end of inflation. Thus, to onset the standard hot Big Bang evolution, a separate reheating phase needs to take place post inflation where the inflaton field transfers its energy to a radiation bath \cite{Bassett:2005xm}. We will refer to this standard picture of inflation  that  requires a separate distinct reheating phase as Cold Inflation (CI) henceforth. Though well-supported by the current observations, this simple picture of inflation raises a  number of concerns: (i) The inflaton potential needs to be very flat in order to sustain slow roll and result in a quasi-exponential expansion of space. Moreover, in most models of inflation, the inflaton potential, in addition to having a flat wing, should also  feature a minimum to which the inflaton field rolls down post inflation and oscillates in order to reheat the universe. These two criteria severely restrict the form of the inflaton potential in the standard CI picture, making it a  challenging task to realize slow-roll inflation within viable fundamental physics framework. (ii) Though several mechanisms of reheating have been explored in the literature \cite{Bassett:2005xm}, they leave few observational signatures making it  exceedingly difficult to probe the dynamics of reheating observationally.

Warm Inflation (WI) \cite{Berera:1995ie} is a promising alternative inflationary paradigm in which  the inflaton field continuously dissipates its energy to a persistent, yet subdominant, radiation bath (for recent reviews on WI, see \cite{Kamali:2023lzq, Berera:2023liv}). Due to this dissipation of energy the subdominant radiation bath coexists with the inflaton field during WI, and when the universe gracefully exists from WI it smoothly transits to a radiation dominated epoch without any need of invoking a distinct reheating phase post inflation. Therefore, WI naturally alleviates the issue of reheating that arises in the context of CI. Warm Inflation has several other attractive features over CI. For instance\,: (i)  WI generally yields a smaller tensor-to-scalar ratio thereby making room for those potentials which are ruled out in CI. A prominent example being the quartic self-coupling inflaton potential ($\lambda\phi^4$) which  produces large amplitude tensor modes in CI and is ruled out by CMB observations \cite{Bartrum:2013fia}. (ii) In the strong dissipative regime, WI prefers sub-Planckian field displacement during inflation \cite{Bastero-Gil:2019gao, Kamali:2019xnt, Das:2020xmh}, which favors small-field inflation models thereby allowing us to avoid working with large-field inflationary models which are difficult to realize as low-energy effective field theories \cite{Planck:2018jri}. (iii) It was shown that other CI pathologies that arise in supergravity models, like the gravitino problem or the eta-problem, can be easily alleviated in a WI setup \cite{BuenoSanchez:2010ygd, Bartrum:2012tg, Bastero-Gil:2019gao}. Moreover, WI models with quadratic \cite{Berera:2025vsu} or $\alpha$-attractor potentials \cite{Chakraborty:2025jof} are shown to be compatible with the recent ACT results \cite{ACT:2025fju, ACT:2025tim} whereas non-trivial modifications are required to be made to CI models to make them compatible with CMB data \cite{Kallosh:2025ijd}. Smaller tensor-to-scalar ratio \cite{Kamali:2023lzq, Berera:2023liv}), distinct running and running of running of scalar spectral index \cite{Das:2022ubr} as well as unique non-Gaussian features \cite{Bastero-Gil:2014raa, Mirbabayi:2022cbt} can observationally distinguish WI from the standard CI models. 

Another novel feature of WI is that the dissipation of the inflaton field during WI  provides additional friction (aside from the friction arising due to the Hubble expansion) in the inflaton's equation of motion. In the strong dissipative regime this term can dominate and, remarkably, sustain slow-roll of the inflaton field even in very steep potentials. One such example of WI taking place in a very steep part of the inflaton potential can be found in \cite{Das:2020xmh}. This very feature of WI helps WI circumvent the de Sitter Swampland conjecture arising in String Theory \cite{Obied:2018sgi, Garg:2018reu, Ooguri:2018wrx} and  incorporates steep potentials into more viable inflationary models in UV-complete gravity theories than the flatter potentials used in standard CI \cite{Das:2018hqy, Motaharfar:2018zyb, Das:2018rpg, Das:2019hto}.  Therefore, WI also alleviates the restrictions on the form of the inflaton potential imposed by CI and can accommodate inflaton potentials of larger variety than CI. 

In this paper we explore WI in the  Witten-O'Raifeartaigh potential \cite{Witten:1981kv, ORaifeartaigh:1975nky}, which features a  steep left wing and a flatter right wing. The Witten-O'Raifeartaigh potential \cite{Witten:1981kv, ORaifeartaigh:1975nky}   belongs to a class of theoretically well-motivated scalar potentials in supersymmetric field theories that implement spontaneous F-term supersymmetry breaking. CI has previously been explored in the flatter right wing of the Witten-O'Raifeartaigh potential \cite{Albrecht:1983ib} where  slow roll of the inflaton field is ensured by the standard background Hubble friction. It was shown that such a CI model with the Witten-O'Raifeartaigh potential is in good agreement with the CMB observations \cite{Martin:2013tda}. The primary focus of this work is to demonstrate that WI can successfully take place on the very steep left wing of this potential  in the strong dissipative regime where the dissipative friction term dominates the WI dynamics, and that this model, too, is in good agreement with current observations.

However, another interesting possibility arises post inflation when the residual kinetic energy of the inflaton field helps it climb up the shallow right wing of the Witten-O'Raifeartaigh potential. In such a case, there can arise a second accelerating phase when the field rolls back  towards the minimum of $V(\phi)$ along the flat right wing of the potential. This second accelerating phase can represent the current accelerating Dark Energy epoch of the Universe, with the same inflaton field currently playing the role of quintessence. Therefore, one can unify inflation and quintessence within a single setup. Such models, where (cold) inflation and quintessence can be realized by the dynamics of a single scalar field, are generally referred to as the quintessential inflaton models, see for instance \cite{Peebles:1998qn, Peloso:1999dm, Dimopoulos:2000md, Majumdar:2001mm, Sami:2004xk, Dimopoulos:2001ix, Giovannini:2003jw, Rosenfeld:2005mt, Hossain:2014xha, Dimopoulos:2017zvq, Rubio:2017gty, Haro:2019gsv, Kleidis:2019ywv, Benisty:2020qta, Benisty:2020xqm, Jaman:2022bho, Mishra:2024qhc} and references therein. However, such unified models, where inflation takes place in the conventional cold inflation setup, usually face two hurdles. Firstly, the inflaton field in any quintessential inflation model needs to be sufficiently stable so as to survive the entire 13 billion years of its post-inflationary history and act as quintessence closer to the present epoch. This requirement severely inhibits the inflaton from  decaying into other degrees of freedom during the reheating process which in most CI models succeeds inflation. It is therefore desirable to invoke  alternative mechanisms, in which reheating does not involve inflaton decay. These include  gravitational reheating \cite{Ford:1986sy, Chun:2009yu},   instant preheating \cite{Felder:1998vq, Campos:2002yk}, curvaton reheating \cite{Feng:2002nb, BuenoSanchez:2007jxm}, non-minimal \cite{Dimopoulos:2018wfg} or Ricci reheating \cite{Opferkuch:2019zbd}, etc. Secondly, the energy scales of these two accelerated epochs differ by many orders of magnitudes. While the early inflationary epoch  is often assumed to take place closer to the GUT energy scale ($10^{16}$ GeV) \cite{Liddle:1993ch}, the current dark energy epoch is associated with  a substantially lower energy scale $\sim 10^{-47}$ GeV$^4$ \cite{Steinhardt:2006bf}. In their original quintessential inflation paper \cite{Peebles:1998qn}, Peebles and Vilenkin addressed this problem by suggesting two different forms of scalar potentials (with different normalizations) for describing these two widely separated accelerating epochs. 

But, as WI doesn't call for a separate reheating phase, it does not face issues relating to reheating which occur in generic CI based quintessential inflation models. Quintessential inflation  has previously been explored a few times in the WI setup \cite{Dimopoulos:2019gpz, Rosa:2019jci, Lima:2019yyv}. In the first two papers \cite{Dimopoulos:2019gpz, Rosa:2019jci}, the authors considered two separate potentials of the quintessential field for the early- and late-time evolution, just like in standard quintessential inflation models. Moreover, dissipation of the scalar field, which is a signature of WI, was considered to be effective only during the inflationary period in these models, whereas the quintessence dynamics was treated in the standard non-dissipative fashion. Later in \cite{Lima:2019yyv}, warm quintessential inflation was achieved with a single scalar potential of the form $V(\phi)=V_0\exp(-\alpha\phi^n/m_p^n) \,(n\geq2),$ (such a potential was first explored in the context of cold quintessential inflation in \cite{Geng:2015fla}). Moreover this paper suggested that the scalar field dissipates its energy into both radiation and matter degrees of freedom. It dissipates more to radiation during inflation (yielding WI) which ceases once inflation ends. Thereafter the scalar field continues to dissipate to matter during the later stages of evolution. Such a construction not only explains inflation and dark energy, but also produces dark matter through dissipation of the scalar field. Thus, this model attempts to explain inflation, dark matter and dark energy within one single framework. The early and late-time stability of this unified model was analyzed in \cite{Das:2023rat}. Baryogenesis was also explored in a similar warm quintessential inflation model in \cite{Basak:2021cgk}. It is worth pointing out that in all these models of warm quintessential inflation, very flat potentials were chosen for the dynamics of the inflaton and the quintessence field, allowing slow-roll dynamics to take place.

In this paper too, we will explore the possibility of accommodating quintessential inflation dynamics as we have two wings of Witten-O'Raifeartaigh potential at our disposal for slow-roll dynamics.  However, as we will demonstrate, in order to successfully accommodate both inflation and quintessence with this single potential, one needs to engineer a few additional features into the present model, such as (i) one needs to have  two different normalizations for  the two wings of the Witten-O'Raifeartaigh potential in order to account for the  extreme hierarchy of energy scales between inflation and the Dark energy dominated acceleration, and (ii) facilitating the slow roll of the quintessence field in the right wing of the Witten-O'Raifeartaigh potential, too, requires additional dissipative friction, just like the inflaton field in WI in steep potentials, in order to maintain  slow roll. 

We have organized the rest of the paper as follows. In Sec.~\ref{WR-brief} we will briefly discuss the Witten-O'Raifeartaigh potential as a successful inflationary model in CI. In Sec.~\ref{WI-brief} we will  provide a brief overview of the standard dynamics of WI and will discuss a suitable choice of the dissipative coefficient for our warm quintessential inflation model. In Sec.~\ref{graceful-exit}, we will determine the condition of gracefully exiting from WI in our model as this is essential for explaining the evolution of the later stages. A thorough MCMC analysis of the  WI dynamics in this model will be presented in Sec.~\ref{WI-left}. We will explore the quintessence dynamics in the right wing of the Witten-O'Raifeartaigh potential in Sec.~\ref{quintessence-right} and will show that with a few adjustments, transient Dark Energy can also be successfully realized in our model. Finally, we will discuss and summarize the  key findings of our analysis in the concluding section~\ref{conclusion}.

\section{A Brief Discussion on Witten-O'Raifeartaigh (Cold) Inflation}
\label{WR-brief}

Witten-O'Raifeartaigh Inflation (WRI) deals with an inflaton potential of the form $V(\phi)\propto \ln^2(\phi/\phi_0)$ \cite{Martin:2013tda}. Such a potential was first explored in a new inflationary scenario \cite{Albrecht:1983ib}. This type of potential arises in supersymmetric theories, which was first analyzed  in \cite{Witten:1981kv} as an attempt to address the gauge hierarchy problem. In the original proposal \cite{Witten:1981kv}, a large hierarchy could not be achieved despite the fact that supersymmetry is broken in such models \cite{ORaifeartaigh:1975nky}. Later, these types of potential were reconsidered in the context of SU(5) GUTs, where the hierarchy problem was successfully addressed \cite{Witten:1981nf, Dimopoulos:1982gm}. Such potentials were also explored in supergravity theories \cite{Papantonopoulos:1986gc, Pollock:1987qc}. A concise discussion  on this class  of particle physics models was presented in \cite{Martin:2013tda}.  

\begin{center}
\begin{figure}[h!]
  \includegraphics[width=7.5cm]{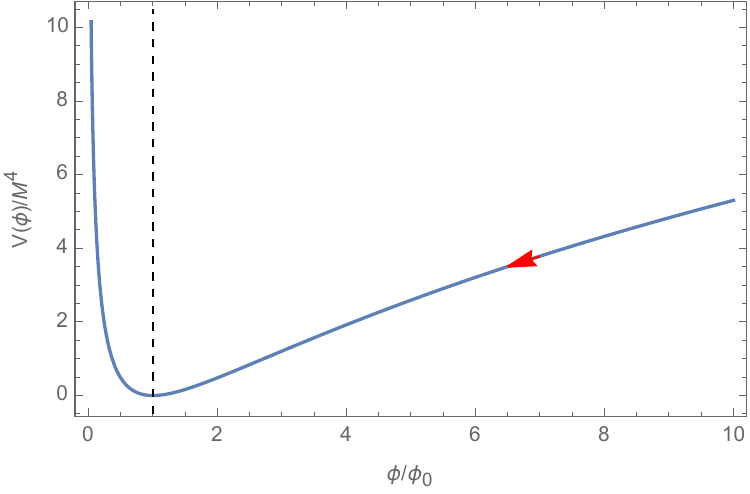}
  \caption
 {Witten-O'Raifeartaigh Inflation potential as given in Eq.~(\ref{WRI-potential}). The dotted line divides the very steep left wing of the potential from its flatter right wing. The arrow  on the right wing depicts the direction along which the slowly rolling inflaton evolves in Cold inflation.}
  \label{WRI-pot}
\end{figure}
\end{center}

Following \cite{Martin:2013tda}, we write the WRI potential as 
\begin{eqnarray}
V(\phi)=M^4\ln^2\left(\frac{\phi}{\phi_0}\right),
\label{WRI-potential}
\end{eqnarray}
where $\phi_0$ is a free parameter whose  typical prior is $\phi_0 \sim \mathcal{O}\left(m_{p}\right)$ (with $m_p = 1/\sqrt{8\pi G}$ being the reduced Planck mass in natural units). The potential is depicted in Fig.~\ref{WRI-pot}. This potential has a very steep left wing and a flatter right wing. It is well known that the left wing of this potential is too steep to sustain CI. To illustrate  this, we look at the first Hubble slow-roll parameter for this potential in CI \cite{Martin:2013tda}:
\begin{eqnarray}
\epsilon_H=-\frac{\dot H}{H^2}\simeq \frac{m_{p}^2}{2}\left(\frac{V_{,\phi}}{V}\right)^2=\frac{2m_{p}^2}{\phi_0^2}\frac{1}{x^2\ln^2x},
\label{epsH-CI}
\end{eqnarray}
where $x\equiv\phi/\phi_0$. In the steeper left branch, as $x<1$, we will always get $\epsilon_H>1$ with the prior $\phi_0=m_{p}$. Unless, one considers $\phi_0\gg m_{p}$, which is, on the other hand, is difficult to justify in the particle physics constructions of such a model, CI cannot take place in this steep left branch of the potential \cite{Martin:2013tda}. 

Thus the slow-roll inflationary predictions were analyzed only for the right branch of this potential in \cite{Martin:2013tda}. The arrow in Fig.~\ref{WRI-pot} indicates the  direction along which slow roll of the inflaton proceeds in CI. Observational predictions of this model in the scalar spectral index\,---\,tensor-to-scalar ratio plane ($n_s$\,---\,$r$ plane) are depicted in Fig.~156 of \cite{Martin:2013tda}. Witten-O'Raifeartaigh potential has also been explored in CI with non-minimally coupled inflaton field \cite{Santos:2022aeb, Santos:2023hhk}. However, here too, the inflation takes place in the flatter right wing of the potential. 

However, as discussed in the introduction of the paper, WI can take place in the steep left wing of this potential in the strong dissipative regime,  where dissipation of the inflaton is significant, yielding large enough friction in its dynamics, thereby sustaining a slow-roll evolution despite the ultra-steep nature of the left wing. We move on to provide a brief discussion of the basic dynamics of WI inflation next.


\section{Essentials of generic Warm Inflation dynamics}
\label{WI-brief}

In WI, the inflaton field keeps dissipating its energy to a radiation bath throughout  the inflationary phase. This helps maintain a subdominant yet non-negligible, nearly constant radiation energy density, $\rho_r$, during inflation despite the near-exponential background expansion. It is generally assumed that the radiation bath maintains thermal equilibrium  so that a temperature $T$ can be assigned to the bath.  Dissipation of the inflaton field  generates an additional friction term (apart from the usual Hubble friction term present in the  Klein-Gordon equation) in inflaton's equation of motion as\,:
\begin{eqnarray}
\ddot\phi+3H\dot\phi+V_{,\phi}=-\Upsilon_r(\phi,T)\dot\phi,
\end{eqnarray}
where $\Upsilon_r(\phi,T)$ is the dissipative coefficient that depends on the inflaton field as well as the on the temperature of the radiation bath. Therefore, depending on which of these two friction terms dominates the inflaton's dynamics, WI can be categorized into two regimes: weak dissipative regime and strong dissipative regime. Defining a dimensionless quantity $Q$ as the ratio of these two friction terms:
\begin{eqnarray}
Q\equiv \frac{\Upsilon_r}{3H},
\end{eqnarray}
weak dissipative regime is the one when the Hubble friction dominates the inflaton's motion (as in the case of CI) and thus $Q \ll 1$, whereas in strong dissipative regime $Q\gg1$ indicating that the dissipative friction term governs the inflaton's evolution. The dissipative term $\Upsilon_r$, in general, takes a simple form \cite{Kamali:2023lzq}:
\begin{eqnarray}
\Upsilon_r(\phi,T)=C_\Upsilon T^p\phi^cM^{1-p-c},
\label{upsilon}
\end{eqnarray}
where $C_\Upsilon$ is a dimensionless quantity whereas $M$ is some appropriate mass scale. Moreover, $p$ and $c$ are integer constants. 

The full set of background evolution equations in WI can be written as \cite{Kamali:2023lzq}
\begin{eqnarray}
&&\ddot\phi+3H(1+Q)\dot\phi+V_{,\phi}=0,\\
&&\dot\rho_r+4H\rho_r=\Upsilon_r(\phi,T)\dot\phi^2,\\
&&H^2=\frac{1}{3m_p^2}\left(\frac{\dot\phi^2}{2}+V(\phi)+\rho_r\right),
\end{eqnarray}
where the first equation is the equivalent of the standard Klein-Gordon equation, the second equation determines the evolution of the subdominant radiation bath, and the third equation is the Friedmann equation. Unlike CI, WI can sustain slow-roll in very steep potentials due to the extra friction provided by dissipation. One can observe from Eq.~(\ref{epsH-CI}) that in slow-roll CI $\epsilon_H\sim \epsilon_V$, where $\epsilon_V$ is the standard potential slow-roll parameter 
\begin{eqnarray}
\epsilon_V=\frac{m_p^2}{2}\left(\frac{V_{,\phi}}{V}\right)^2.
\end{eqnarray}
Thus, for steep potentials where $\epsilon_V>1$, one gets $\epsilon_H>1$ which violates the slow-roll condition. In contrast, in WI, it turns out that 
\begin{eqnarray}
\epsilon_H\sim \frac{\epsilon_V}{(1+Q)}. 
\label{epsH-WI}
\end{eqnarray}
This relation can be easily derived from the slow-roll approximated evolution equations given as 
\begin{eqnarray}
&&3H(1+Q)\dot\phi\approx -V_{,\phi},\\
&&H^2\approx \frac{V}{3m_p^2}.
\end{eqnarray}
Therefore, in a strong dissipative regime where $Q\gg1$, one can have $\epsilon_V>1$ (i.e. steep potentials) without violating the slow-roll condition $\epsilon_H<1$. 

It has been shown that WI driven by an axionic field coupled to non-Abelian gauge fields successfully takes place  in the strong dissipative regime, and such a model was dubbed Minimal Warm Inflation (MWI) \cite{Berghaus:2019whh}. The dissipative coefficient $\Upsilon_r$ of the axionic inflaton field was shown to be of cubic power of $T$ without any dependence on $\phi$. Thus, according to the form of the dissipative coefficient given in Eq.~(\ref{upsilon}), for MWI  $p=3$ and $c=0$. In our present model we will also consider a similar dissipative coefficient for the inflaton field phenomenologically, without claiming any axionic nature of the quintessential inflaton field, and will show that WI takes place  in the strong dissipative regime in our model.

\begin{center}
\begin{figure}[h!]
  \includegraphics[width=8.5cm]{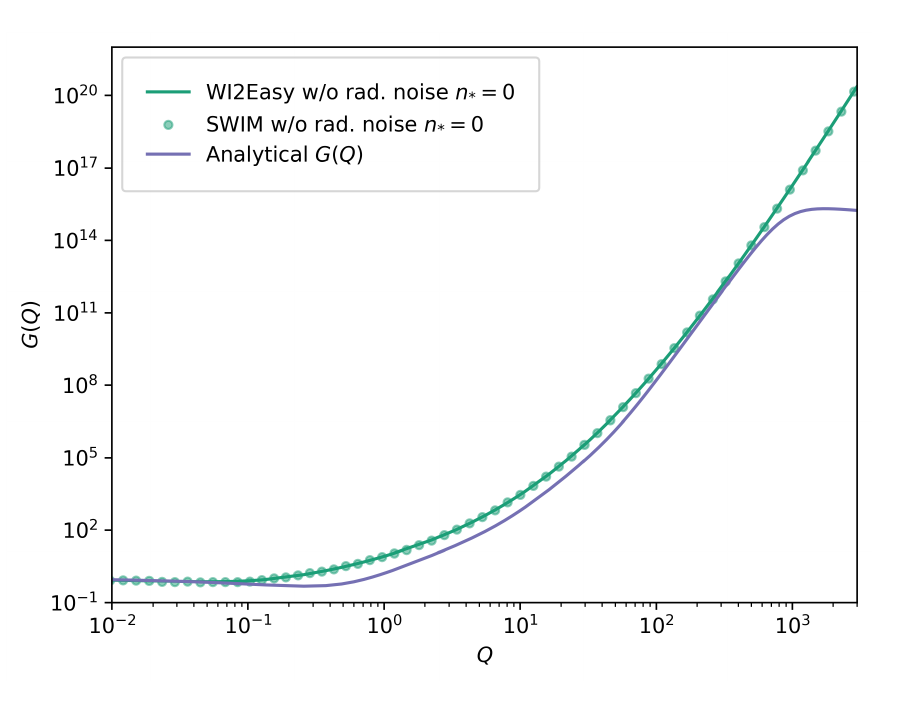}
  \caption
 {The comparison of the approximated analytical form of G(Q) proposed in \cite{Das:2020xmh} with the numerically generated ones using \texttt{WI2easy} \cite{Rodrigues:2025neh} and \texttt{SWIM} \cite{Umang}}
  \label{GQ-comparison}
\end{figure}
\end{center}

Since the inflaton field is coupled to the radiation bath, the primordial scalar power spectrum in WI, under slow-roll approximations,  takes the form \cite{Ramos:2013nsa, Benetti:2016jhf}
\begin{eqnarray}
{\cal P}_{\mathcal R}={\cal P}_{\mathcal R}^{\rm CI} \times {\mathcal F}(Q),
\label{eq:PS_WI}
\end{eqnarray}
where ${\cal P}_{\mathcal R}^{\rm CI}$ is the  power spectrum of comoving curvature fluctuations in CI under slow-roll approximations, which is given by
\begin{eqnarray}
{\cal P}_{\mathcal R}^{\rm CI} = \left(\frac{H^2}{2\pi\dot{\phi}}\right)^2,
\label{eq:PS_CI}
\end{eqnarray}
and the function ${\mathcal F}(Q)$ is given by
\begin{eqnarray}
{\mathcal F}(Q)\equiv \left(1 + 2  n_{\rm BE}(T) +\frac{2\sqrt3\pi Q}{\sqrt{3 + 4 \pi Q}}\frac{T}{H}\right)G(Q).
\end{eqnarray}
Here $n_{\rm BE}(T)$ is the occupation number corresponding to the thermal (Bose-Einstein) distribution  of the inflaton field when and if the inflaton field thermalises with the coexisting thermal bath. It has been observed that in strong dissipation, the scalar power depends very mildly on whether or not the inflaton field themalizes during WI. The growth factor, $G(Q)$, takes care of the coupling of the inflaton and radiation fluctuations. In general, $G(Q)$ can only be determined numerically by solving the set of perturbation equations in WI. The specific form of $G(Q)$ depends mostly on the form of the dissipation function $\Upsilon_r$ and has mild dependence on the form of the inflaton potential as far as strong dissipative models are concerned. An approximated analytic form of $G(Q)$ for MWI with runaway exponential potential was proposed in \cite{Das:2020xmh} which can be written as 
\begin{eqnarray}
G(Q)&=&\frac{1+6.12 \, Q^{2.73}}{(1 + 6.96 \, Q^{0.78})^{0.72}}\nonumber\\
&&+\frac{0.01 \, Q^{4.61}(1+4.82\times 10^{-6} \, Q^{3.12})}{(1+6.83\times 10^{-13} \, Q^{4.12})^2}.
\end{eqnarray}
As the $G(Q)$ depends very weakly on the form of the potential, one can, in principle, use this form of $G(Q)$ even in the present case under consideration. However, at least two numerical codes, namely \texttt{WarmSPy} \cite{Montefalcone:2023pvh} and \texttt{WI2easy} \cite{Rodrigues:2025neh}, are now available in the literature that provide numerical forms of the function $G(Q)$ as a function of $Q$. A new code, \texttt{SWIM} (Stochastic Warm Inflation Module) \cite{Umang}, is also being developed to further test the forms $G(Q)$ produced by the existing codes. We make use of \texttt{WI2easy} \cite{Rodrigues:2025neh} and \texttt{SWIM} \cite{Umang} to generate the numerical form of the $G(Q)$ function, and have compared it with the above approximated analytical form. We furnish this comparison in Fig.~\ref{GQ-comparison}, and note that the analytical form approximately follows the numerically generated $G(Q)$, but deviates significantly as $Q$ becomes larger than $10^3$. As we will probe our WI model in strong dissipative regime, it would thus be wise to use the numerical form of $G(Q)$ generated by \texttt{WI2easy} \cite{Rodrigues:2025neh} and \texttt{SWIM} \cite{Umang} for analysis of the model. 

The amplitude of the scalar power spectrum is fixed at a specific pivot scale, $k_*$, at a time when this scale crosses the horizon during inflation. 
The scalar spectral index $n_s$ and its running $\alpha_s$ can be determined at the horizon crossing of the pivot scale $k_*$ as
\begin{eqnarray}
n_s-1&=&\left.\frac{d\ln {\mathcal P}_{\mathcal R}(k/k_*)}{d\ln(k/k_*)}\right|_{k\rightarrow k_*},\\
\alpha_s&=&\left.\frac{dn_s(k/k_*)}{d\ln(k/k_*)}\right|_{k\rightarrow k_*}
\end{eqnarray}

On the other hand, as the tensor perturbations in WI do not couple to the radiation bath, the primordial tensor spectrum takes the same form in WI as in CI:
\begin{eqnarray}
{\mathcal P}_T={\mathcal P}_T^{\rm CI}=\frac{H^2}{2\pi^2m_p^2}.
\end{eqnarray}
The tensor-to-scalar ratio thus can be determined as 
\begin{eqnarray}
r=\left.\frac{{\mathcal P}_T(k/k_*)}{{\mathcal P}_{\mathcal R}(k/k_*)}\right|_{k\rightarrow k_*}.
\end{eqnarray}
These background equations and the form of the power spectra will help analyse the Witten-O'Raifeartaigh Warm Inflation (WRWI) taking place in the steep left wing of the potential. However, before we analyse the model, it is essential to figure out whether WI can make a graceful exit from slow roll in such a steep wing of the potential. The next section is dedicated to  analyzing the graceful-exit condition for  this model.


\section{Analysis of Graceful exit condition when Warm Inflation takes place in the left wing of Witten-O'Raifeartaigh potential}
\label{graceful-exit}

We need to first adjudge whether WI will gracefully exit if inflation takes place in the steep part of the potential. To investigate this, we will follow the analysis of \cite{Das:2020lut}. We see from Eq.~(\ref{epsH-WI}) that $1+Q$  must be  greater than $\epsilon_V$ initially so that inflation can initiate ($\epsilon_H\sim\epsilon_V/(1+Q)<1$), and WI ends when $\epsilon_V\sim 1+Q$. Now, in WI, both $\epsilon_V$ and $Q$ evolves with time or number of $e$-foldings as inflation progresses. Thus, there can be three different scenarios in which WI can make a graceful exit:
\begin{enumerate}
\item When $\epsilon_V$ increases with $N$: WI ends if $Q$ remains constant or decreases. WI can also end if $Q$ increases with a smaller rate than $\epsilon_V$.
\item When $\epsilon_V$ does not evolve with $N$ (constant $\epsilon_V$): WI only ends if $Q$ decreases with $N$.
\item When $\epsilon_V$ decreases with $N$: In such a case graceful exit demands a faster rate of decrease in $Q$ than $\epsilon_V$. In addition to that, the condition $\epsilon_V=1+Q$ should meet before the value of $\epsilon_V$ decreases below unity, otherwise there will be no graceful exit. The later part needs to be checked numerically. 
\end{enumerate}

It is straightforward to calculate that, with a general dissipative coefficient of the form, given in Eq.~(\ref{upsilon}),
$\epsilon_V$ and $Q$ evolve as \cite{Das:2020lut}
\begin{eqnarray}
\frac{d\ln\epsilon_V}{dN}=\frac{4\epsilon_V-2\eta_V}{1+Q}, \label{eps-evolve}
\end{eqnarray}
and
\begin{eqnarray}
\frac{d\ln Q}{dN}=\frac{(2p + 4)\epsilon_V - 2p \, \eta_V - 4c \, \kappa_V}{C_Q} \, ,
\label{Q-evolve}
\end{eqnarray}
where
\begin{eqnarray}
C_Q\equiv 4 - p+(4+p)Q.
\label{eq:CQ}
\end{eqnarray}
Here $\eta_V$ is the standard second potential slow-roll parameter:
\begin{eqnarray}
\eta_V=m_p^2\left(\frac{V_{,\phi\phi}}{V}\right),
\end{eqnarray}
and $\kappa_V$ is defined as 
\begin{eqnarray}
\kappa_V \equiv m_p^2 \,  \frac{V_{,\phi}}{\phi V}.
\end{eqnarray}

For convenience, we rewrite the WRI potential given in Eq.~(\ref{WRI-potential}) as 
\begin{eqnarray}
V(\phi)=V_0\ln^2\left(\lambda \frac{\phi}{m_p}\right),
\end{eqnarray}
where now $\lambda$ would be the free parameter of the theory. Comparing this form of the potential with the one in Eq.~(\ref{WRI-potential}), we see that $\phi_0\equiv m_p/\lambda$. We can see setting $\lambda\geq 1$ will probe the theories for scale $\phi_0$ smaller than the Planck mass, which is well motivated from an effective field theory prospective. 

\begin{figure}[ht!]
    \centering
    \includegraphics[width=0.48\textwidth]{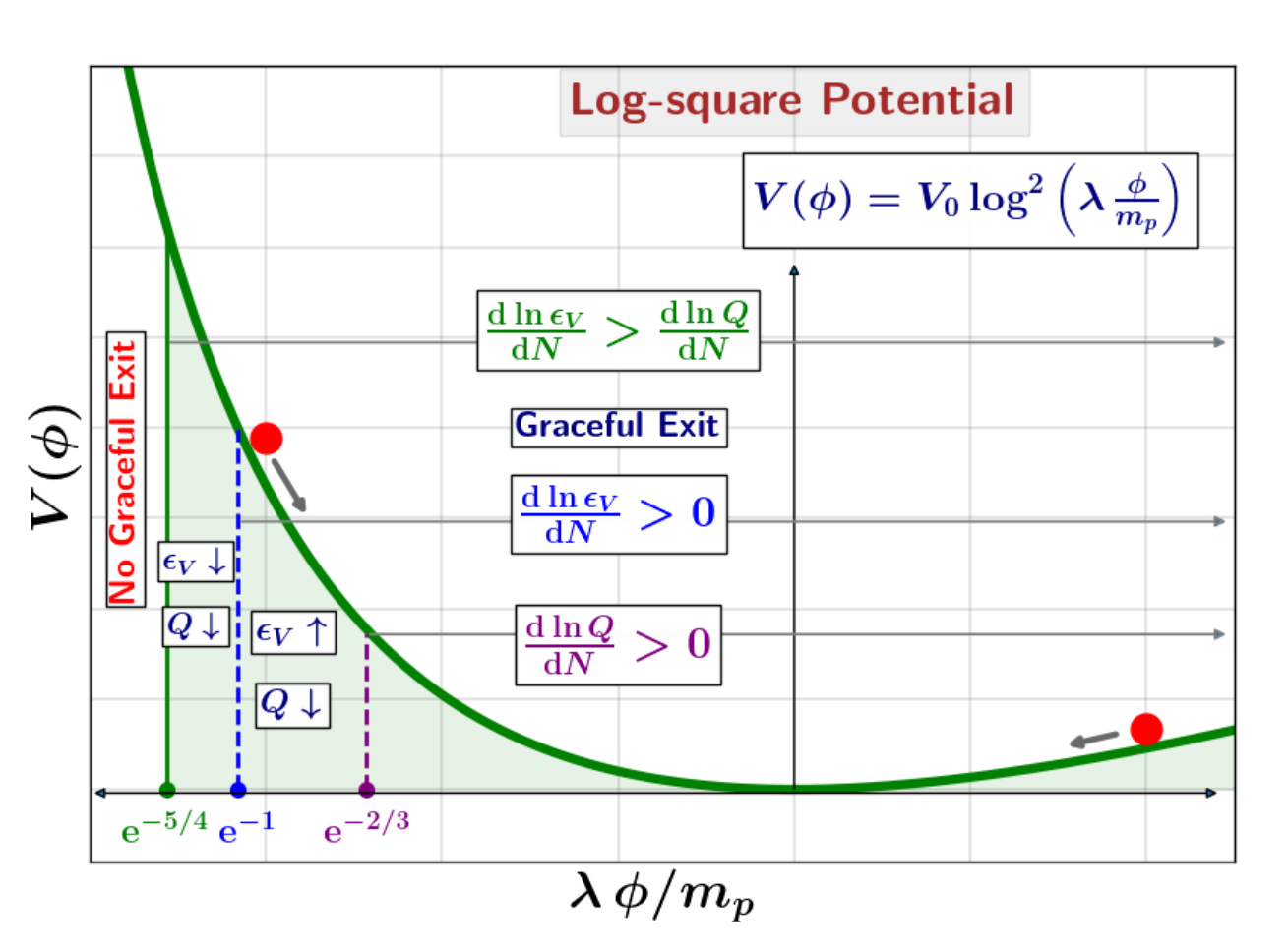}
    \caption{The Witten-O'Raifeartaigh potential plotted schematically, marking the regions of graceful exit from WI.}
    \label{fig:regimes_pot_logsqr}
\end{figure}

With this potential, we calculate $\epsilon_V$, $\eta_V$ and $\kappa_V$  {\violet to be} 
\begin{eqnarray}
    \epsilon_V(\phi) &=& \frac{2m_p^2}{\phi^2} \, \frac{1}{\ln^2(\lambda\phi/m_p)}, \label{eq:epsilon_logsqr} \\
    \eta_V(\phi) &=& \frac{2m_p^2}{\phi^2}\frac{1-\ln(\lambda\phi/m_p)}{\ln^2(\lambda\phi/m_p)}, \label{eq:eta_logsqr}  \\
    \kappa_V(\phi)&=& \frac{2m_p^2}{\phi^2} \, \frac{1}{\ln(\lambda \phi/m_p)} \label{eq:kappa_logsqr},  
\end{eqnarray}
respectively. Inserting these expressions into Eq.~(\ref{eps-evolve}) and Eq.~(\ref{Q-evolve}) yields for $p=3$, $c=0$ (MWI model dissipative coefficient) and $Q \gg 1$ (strong dissipation)
\begin{eqnarray}
    Q \, \frac{d\ln\epsilon_V}{dN}&=&\frac{4m_p^2}{\phi^2\ln^2(\lambda\phi/m_p)}\left[1+\ln\left(\lambda\frac{\phi}{m_p}\right)\right],\label{eq:epsilon_N_logsqr} \\
    7Q \, \frac{d\ln Q}{dN}&=&\frac{4m_p^2}{\phi^2\ln^2(\lambda\phi/m_p)}\left[2+3\ln\left(\lambda\frac{\phi}{m_p}\right)\right] . \label{eq:Q_N_logsqr}
\end{eqnarray}
We see from the above equations that when $\phi/m_p\gtrsim 1/(e\lambda)\sim 0.37/\lambda$, $\epsilon_V$ increases, while it decreases when $\phi/m_p\lesssim 1/(e\lambda)\sim 0.37/\lambda$. However, $Q$ increases when $\phi/m_p\gtrsim 1/( e^{2/3}\lambda)\sim 0.51/\lambda$ and decreases for  $\phi/m_p\lesssim 1/( e^{2/3}\lambda)\sim 0.51/\lambda$.  Therefore, we need to analyze the situation in three regions:
\begin{enumerate}
    \item $\phi/m_p\gtrsim 0.51/\lambda$: Both $\epsilon_V$ and $Q$ are increasing in this region. Thus we need $d\ln Q/dN < d\ln\epsilon_V/dN$ for graceful exit in this region. This condition leads to $\phi/m_p> 0.29/\lambda$ for strong dissipative regime, which is consistent.
    \item $0.37/\lambda \lesssim \phi/m_p\lesssim 0.51/\lambda$: In this region $\epsilon_V$ increases and $Q$ decreases. Therefore, graceful exit is guaranteed. 
    \item $\phi/m_p\lesssim 0.37/\lambda$: Both $\epsilon_V$ and $Q$ are decreasing in this region. Thus we need $|d\ln Q/dN|>|d\epsilon_V/dN|$ for graceful exit in this region. This condition leads to $\phi/m_p> 0.29/\lambda$ for strong dissipative regime. Therefore, there is only a small window $0.29/\lambda<\phi/m_p\lesssim 0.37/\lambda$ where $Q$ is decreasing faster than $\epsilon_V$. 
\end{enumerate}
From the above analysis we note that graceful exit is always guaranteed if $\phi/m_p> 0.37/\lambda$, whereas $\phi/m_p< 0.37/\lambda$ region is largely inconsistent with graceful exit. This suggests that though graceful exit always takes place on the right flatter wing on the potential, on the left steeper wing, it is only guaranteed in the region $\phi/m_p> 0.37/\lambda$. We depict all these regions based on the graceful exit condition in Fig.~\ref{fig:regimes_pot_logsqr}.

\section{Warm Inflation in the steep left wing of the Witten-O'Raifeartaigh potential}
\label{WI-left}

\begin{figure}[ht!]
    \centering
    \includegraphics[width=0.48\textwidth]{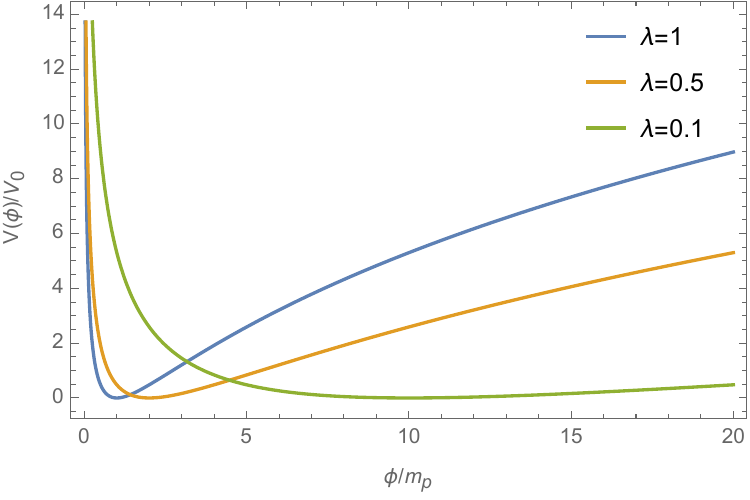}
    \caption{The Witten-O'Raifeartaigh potential for different values of $\lambda$.}
    \label{fig_pot_lambda}
\end{figure}

With the given graceful exit conditions furnished in the previous section, we see that Warm Inflation can take place in the steep left wing of the Witten-O'Raifeartaigh potential, but the last several $e$-folds of inflation will take place near the bottom of the potential as that is the region where Warm Inflation can gracefully exit. Thus, we can consider that Warm Inflation will mainly take place in the left branch but near the bottom of the potential. It is important to note that if we take values of $\lambda$ smaller than unity (which, in turn, will result in super-Planckian values of $\phi_0$ that might be  problematic from an effective field theory point of view), the steep left wing will become shallower with decreasing $\lambda$ near the bottom of the potential, and the left wing will keep becoming less steeper, as has been shown in Fig.~\ref{fig_pot_lambda}. As the potential looses its steepness in its left wing with decreasing values of $\lambda$, Warm Inflation will take place in weak dissipative regime when $\lambda$ becomes lesser than unity. Therefore, in our following analysis,  we will consider  $\lambda\geq1$ for which Warm Inflation takes place in the strong dissipative regime. Additionally, $\lambda\geq1$ results in $\phi_0\leq m_p$,  which is well justified from the effective field theory prospective.

\begin{figure*}[ht!]
    \centering
    \includegraphics[width=0.8\textwidth]{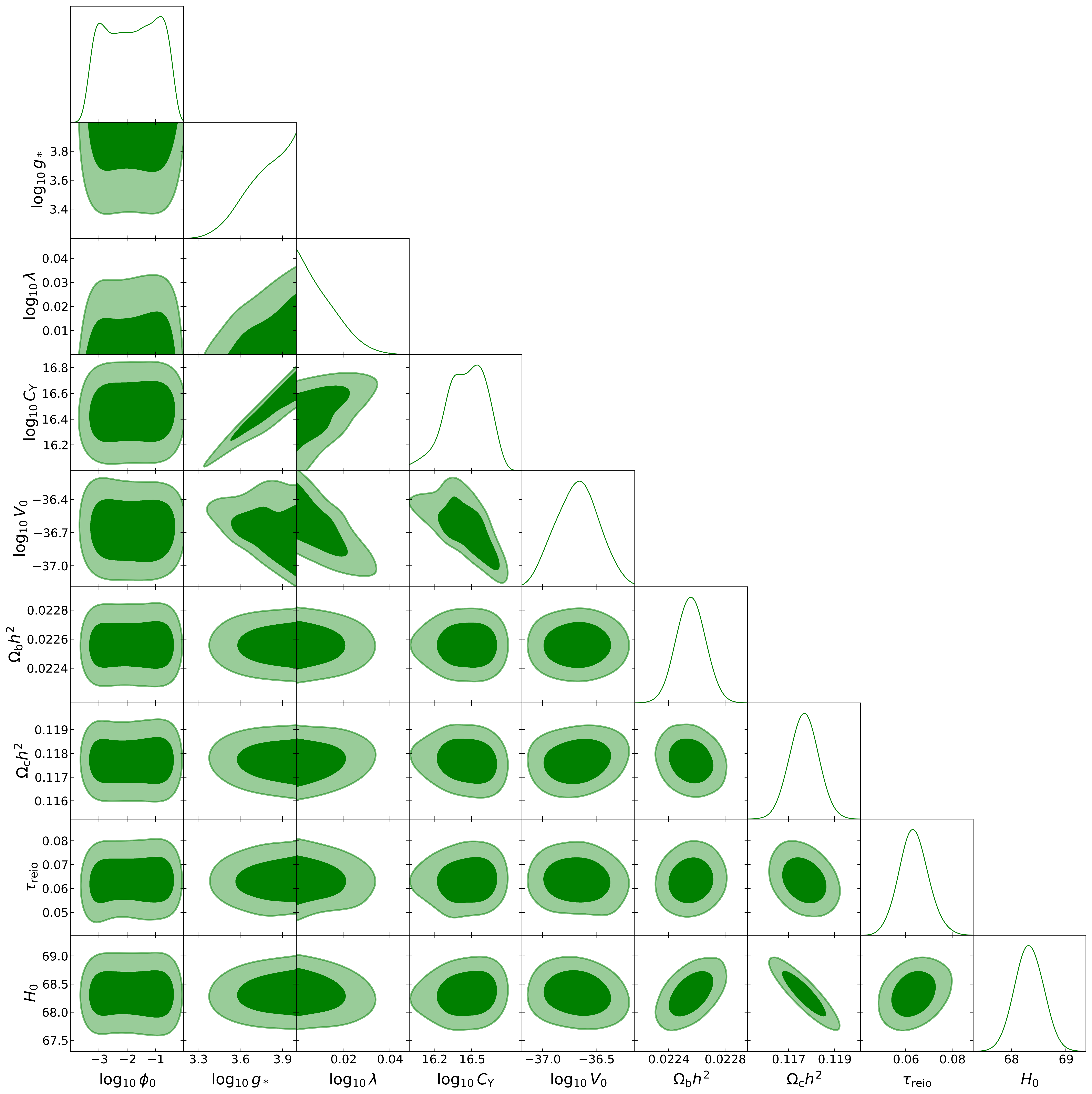}
    \caption{The posterior distribution of parameters of the Warm Inflation model taking place in the left wing of the Witten-O'Raifeartaigh potential}
    \label{fig_contourplot}
\end{figure*}

We have carried out a thorough MCMC analysis of the model by combining the latest cosmological datasets from Planck \cite{Planck:2018jri}, ACT \cite{ACT:2025fju, ACT:2025tim} and DESI \cite{DESI:2024mwx, DESI:2024uvr, DESI:2025zgx} using the publicly available code \texttt{Cobaya}. The methodology to incorporate the Warm Inflationary power spectrum, given in Eq.~(\ref{eq:PS_WI}), into Cobaya (or \texttt{COSMOMC}) by converting the power spectrum into a function of comoving wavenumber $k$ has been developed in \cite{Kumar:2024hju}. We followed this methodology to feed the power spectrum of our model into Cobaya and the results are furnished in Fig.~\ref{fig_contourplot}. It can be seen from Fig.~\ref{fig_contourplot} that $\lambda\sim1.029$ is preferred by the data. The other best-fit model parameters are $g_*\sim6040$, $\tilde C_{\Upsilon}\equiv C_\Upsilon M^{-2}\sim3.18\times 10^{16}\,m_p^{-2}$, and $V_0\sim1.6\times10^{-37}\,m_p^4$. At Hubble-crossing of the pivot scale $\phi_*\sim0.57\,m_p$ and $Q_*\sim967.9$. Warm inflation ends (i.e. when $\epsilon_H\sim1$) at field value $\phi_{\rm end}\sim0.958 \, m_p$ (the minima of the potential  for $\lambda=1.029$ lies at $\phi \simeq 0.972\,m_p$). These values ensure that Warm inflation is taking place in the strong dissipative regime near the bottom part of the left-wing of the potential. Inflation lasts for 42.08 $e$-folds since the Hubble-exit of the pivot scale.  Our analysis yields $A_s\sim2.13\times10^{-9}$, $n_s=0.9762$, $\alpha_s=0.00162$, and $r\sim1.6\times10^{-30}$ (Warm Inflationary models in the strong dissipative regime, in general, produce very low tensor-to-scalar ratio).  All relevant observables are in very good agreement with the latest  datasets emerging from the joint analysis of Planck, ACT and DESI results. We have furnished the best-fit model parameters and the observables of our model in Table~\ref{tab1}.

\begin{widetext}
\begin{center}
\begin{table}[h!]
    \centering
    \begin{tabular}{|c|c|c|c|c|c|c|c|}
    \hline
      \multicolumn{4}{|c|}{Best-fit model parameters}   & \multicolumn{4}{|c|}{Observables} \\
      \hline 
      
      $\lambda$   & $V_0$ & $\tilde C_\Upsilon$ & $g_*$ & $A_s$ & $n_s$ & $r_{0.002}$ & $\alpha_s$\\
      &  (in $m_p^4$) & (in $m_p^{-2}$) & & & & &\\
      \hline
      & & & & & & &\\
      $1.029$  & $1.60\times10^{-37}$ & $3.18\times 10^{16}$ & $6040$ & $2.13\times 10^{-9}$ & $0.9762$ & $1.6\times 10^{-30}$ & $0.00162$ \\
      \hline
    \end{tabular}
    \caption{Best-fit model parameters and observables of WI in our quintessential inflation model from MCMC analysis}
    \label{tab1}
\end{table}
\end{center}
\end{widetext}

\begin{figure}[ht!]
    \centering
    \includegraphics[width=0.48\textwidth]{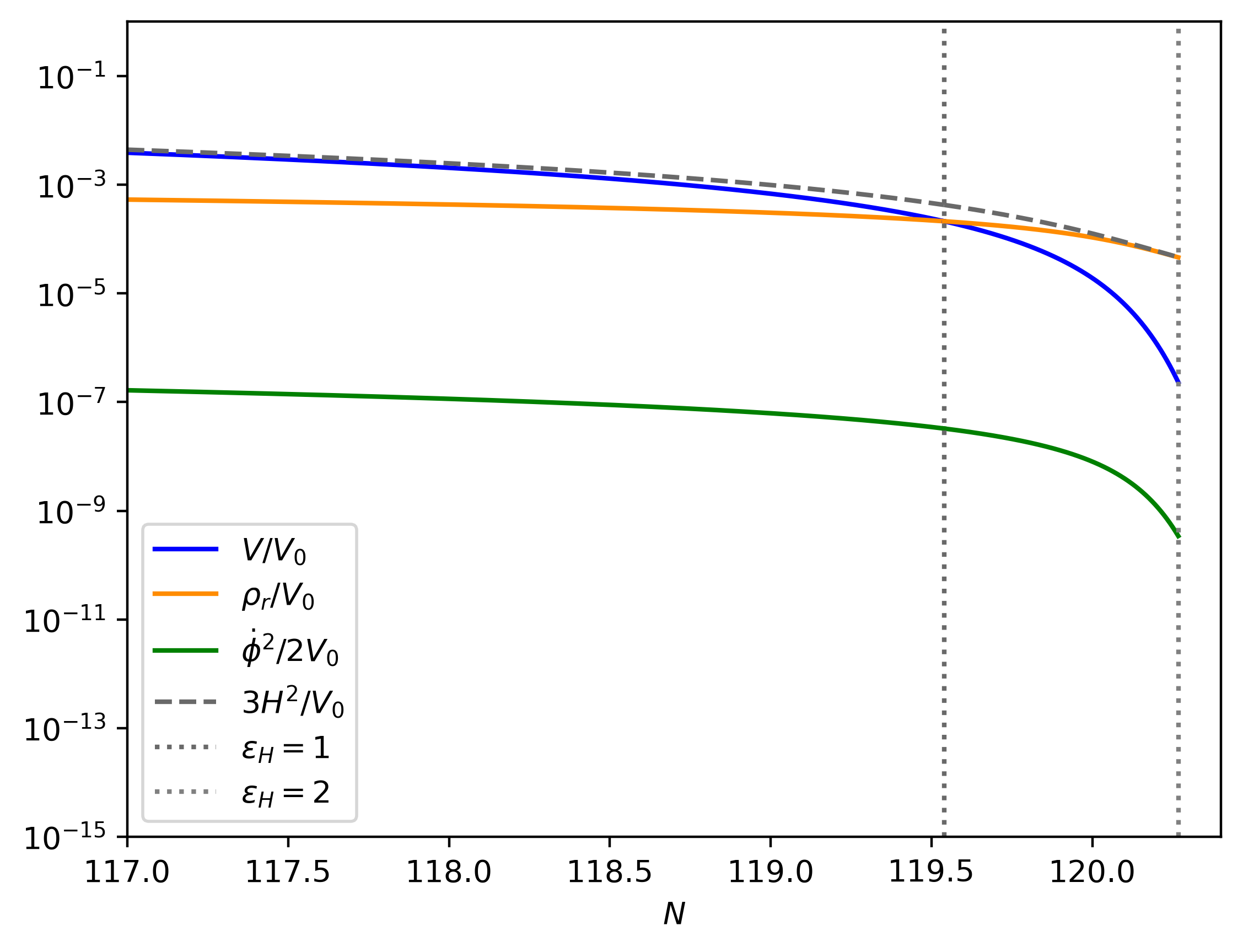}
    \caption{Evolution of potential, kinetic and radiation energy densities near the graceful exit}
    \label{fig_density}
\end{figure}

\begin{figure}[ht!]
    \centering
    \includegraphics[width=0.48\textwidth]{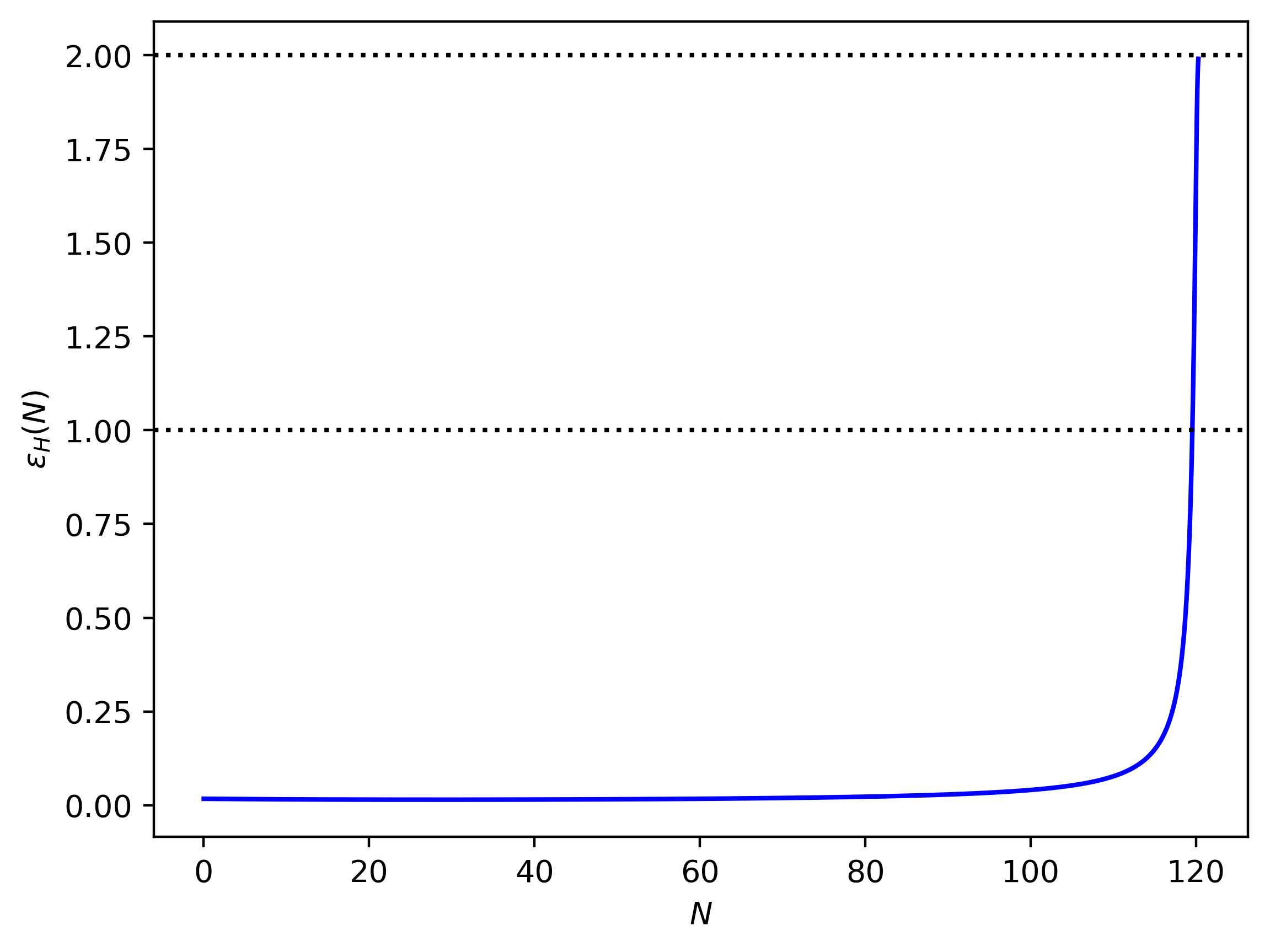}
    \caption{Evolution of the Hubble slow-roll parameter $\epsilon_H$}
    \label{fig_epsH}
\end{figure}

\begin{figure}[ht!]
    \centering
    \includegraphics[width=0.48\textwidth]{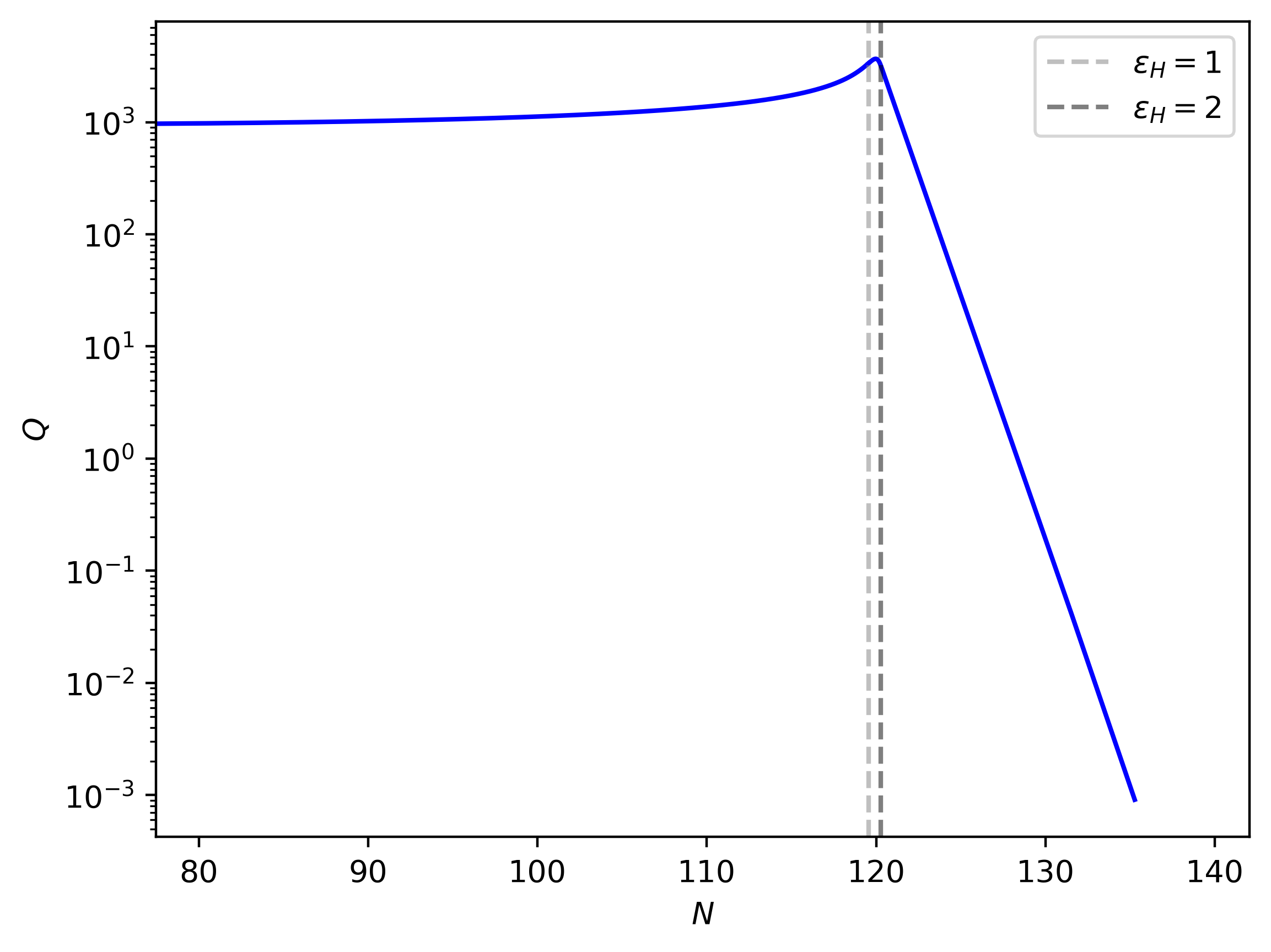}
    \caption{Evolution of the parameter $Q$}
    \label{fig_Q}
\end{figure}

Furthermore, the background analysis of the model with $\lambda\sim1$  yields that the model smoothly transits to a radiation dominated epoch once Warm Inflation gracefully exits. This can be seen from the Fig.~ \ref{fig_density} where we have plotted the kinetic and potential energy densities of the inflaton field as well as the radiation energy density which takes over the inflaton energy density when $\epsilon_H$ reaches 1. It can also be seen that it takes another 0.7 $e$-folds after the  graceful exit to reach a radiation dominated epoch signified by $\epsilon_H\sim2$. In Fig.~\ref{fig_epsH}, we have shown the evolution of the $\epsilon_H$ parameter. We further note from the evolution of the parameter $Q$ in our model, depicted in Fig.~\ref{fig_Q}, that $Q$, and therefore $\Upsilon_r$, steadily decreases after the universe becomes radiation dominated, and after a few $e$-foldings the scalar field will effectively cease to dissipate energy to the radiation bath. A standard radiation dominated epoch will commence thereafter. 

\section{Scope of quintessential Dark Energy in the right wing of the Witten-O'Raifeartaigh potential}
\label{quintessence-right}

After demonstrating that Warm Inflation can successfully takes place on the steep left wing of the Witten-O'Raifeartaigh potential, we will now exploit the flatter right wing to explore the possibility of realizing a second accelerating phase corresponding to the present dark energy epoch. 
We learnt from the previous section that successful inflation on the left steep wing of the the Witten-O'Raifeartaigh potential requires $V_0\sim 1.6\times10^{-37}\,m_p^4$. However dark energy ($\rho_{\rm DE}\sim10^{-120}\,m_p^4$) is many orders of magnitude smaller than $V_0$. Thus, even if the inflaton field rolls up the flat right wing of the Witten-O'Raifeartaigh potential (after sourcing Warm Inflation in the left wing), the scenario still cannot mimic the present Dark Energy epoch even if the inflaton field behaves like quintessence on the right wing. 

In this context it is interesting to note that in the original quintessential inflation model proposed by Peebles and Vilenkin \cite{Peebles:1998qn}, two different forms of the scalar potential were considered for  the inflaton and  quintessence:
\begin{eqnarray}
V(\phi)=\left\{
\begin{array}{ccc}
\tilde\lambda(\phi^4+m^4)&&\phi\leq0\\
\frac{\tilde\lambda m^4}{1+(\phi/m)^{\tilde\alpha}}&& \phi\geq0,
\end{array}
\right.
\label{eq:QI_wings_functions}
\end{eqnarray}
where $\tilde\lambda\sim10^{-15}$ was fixed by the CMB anisotropies, and the value of $\tilde\alpha$ was a tunable parameter that determines the steepness of the quintessence potential. Therefore, it is evident that the normalizations of the two wings of the scalar potential (one giving rise to inflation, and the other providing quintessence) must be different in order to achieve  successful quintessential inflation. In the same spirit, to have successful quintessential inflation with the Witten-O'Raifeartaigh potential, we propose that the normalization $V_0$ must be different in the two wings of the potential, i.e. 
\begin{eqnarray}
V(\phi)=\left\{
\begin{array}{ccc}
V_0\ln^2\left(\lambda\frac{\phi}{m_p}\right)&&\phi\leq m_p/\lambda\\
V_1\ln^2\left(\lambda\frac{\phi}{m_p}\right)&& \phi\geq m_p/\lambda,
\end{array}
\right.
\label{broken-WOR-pot}
\end{eqnarray}
where $V_1$ can now be suitably chosen to determine the steepness of the quintessence potential. In passing, note that the Peebles-Vilenkin potential ~(\ref{eq:QI_wings_functions}) is a $C^2$ curve despite having two different forms of the potential along its two wings. In our case, after choosing two different normalization of the Witten-O'Raifeartaigh potential ~(\ref{broken-WOR-pot}), the curve of the `broken' Witten-O'Raifeartaigh potential becomes a $C^1$ curve. Though, this doesn't hinder the dynamics of the quintessential inflaton field in this model. 

\begin{figure}[ht!]
    \centering
    \includegraphics[width=0.48\textwidth]{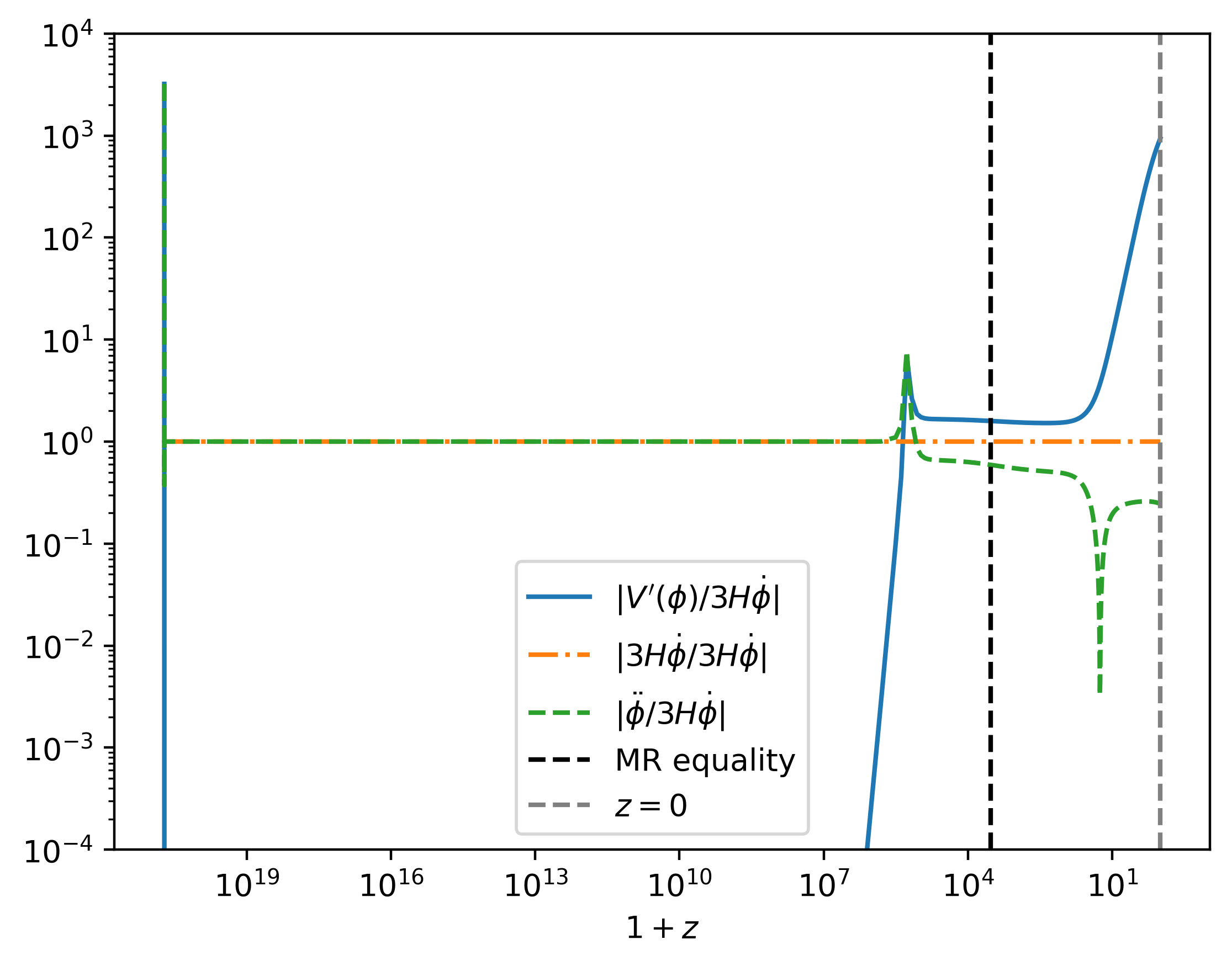}
    \caption{Comparison of the terms in the Klein-Gordon equation of the quintessence field with respect to the Hubble friction for the entire post-inflationary evolution. The blue line shows that the Hubble friction is orders of magnitude larger than the $V'$ term for the most part of the evolution, and it starts to dominate the dynamics of the quintessence field only relatively recently at $z \sim 100$}.
    \label{fig_freezing}
\end{figure}

Moreover, from the analysis of the previous section, we obtained that the field value at the end of ``reheating" (when $\epsilon_H\sim2$) the field value of $\phi_{\rm reh}\sim 0.972\,m_p$, and its kinetic energy is $5.49\times10^{-47}\,m_p^4$. With this kinetic energy the scalar field is not able to climb up the right wing of the potential sufficiently and it freezes at $\phi_q\sim1.147\,m_p$ due to Hubble friction. Once Hubble friction diminishes significantly and the potential energy density of the quintessential field starts to dominate, the quintessence field thaws and begins rolling back towards the bottom of the potential along the right wing. A comparison of the terms in the Klein-Gordon equation of  the quintessence field (post inflation) with the Hubble friction, depicted in Fig.~\ref{fig_freezing}, ensures freezing of the quintessential inflaton field for the entire post-inflationary evolution and shows that the thawing only starts relatively recently at $z \sim 100$. 

However, at $\phi_q$, the potential is still very steep $(\epsilon_V\sim 56)$, and so the field cannot slow-roll back to the minimum {\it slowly} thereby precluding  quintessence-like evolution. Hence, a warm inflation-like dissipative term which operates in the right wing as well as the left wing, helps the quintessence field slow-roll down the flat right wing of the potential and play the role of dark energy.  Consequently, we introduce the dissipative term $\Upsilon_m$ (following \cite{Lima:2019yyv}),
\begin{eqnarray}
\Upsilon_m=\frac{C_m}{\rho_m^{1/4}},
\end{eqnarray}
into the dynamics of the quintessential inflaton field.
As a result the quintessential inflaton field in our WI model will have the full dissipative term 
\begin{eqnarray}
\Upsilon_{\rm total}=\Upsilon_r+\Upsilon_m=C_\Upsilon\frac{T^3}{M^2}+\frac{C_m}{\rho_m^{1/4}},
\end{eqnarray}
which needs to arise from the underlying particle physics picture of the model. The first term $\Upsilon_r$ on the r.h.s will be the dominant dissipative term ($\Upsilon_r\gg\Upsilon_m$) during WI (as has also been argued in \cite{Lima:2019yyv}). Thus, we can ignore the $\Upsilon_m$ term during the inflationary period, and the previous analysis of WI taking place in the left wing of the Witten-O'Raifeartaigh potential remains unaffected. As argued earlier, this dissipative term becomes ineffective soon after the universe enters the radiative epoch. During matter domination,  the gradual decrease in the matter density $\rho_m$ leads to an increase in the $\Upsilon_m$ dissipative term. The increase of $\Upsilon_{m}\dot\phi$ over $3H\dot\phi$ in the equation of motion (\ref{eq:diss}) helps the quintessence field roll back slowly to the bottom of its potential, thereby resulting in a transient epoch of late time acceleration. The dynamics of the quintessential inflaton field during these late epochs is therefore determined by the following set of equations:

\begin{eqnarray}\label{eq:diss}
&&\ddot\phi+3H\dot\phi+\Upsilon_{m}\dot\phi+V_{,\phi}=0,\\
&&\dot\rho_m+3H\rho_m=\Upsilon_m\dot\phi^2.
\end{eqnarray}

\begin{figure}[ht!]
    \centering
    \includegraphics[width=0.48\textwidth]{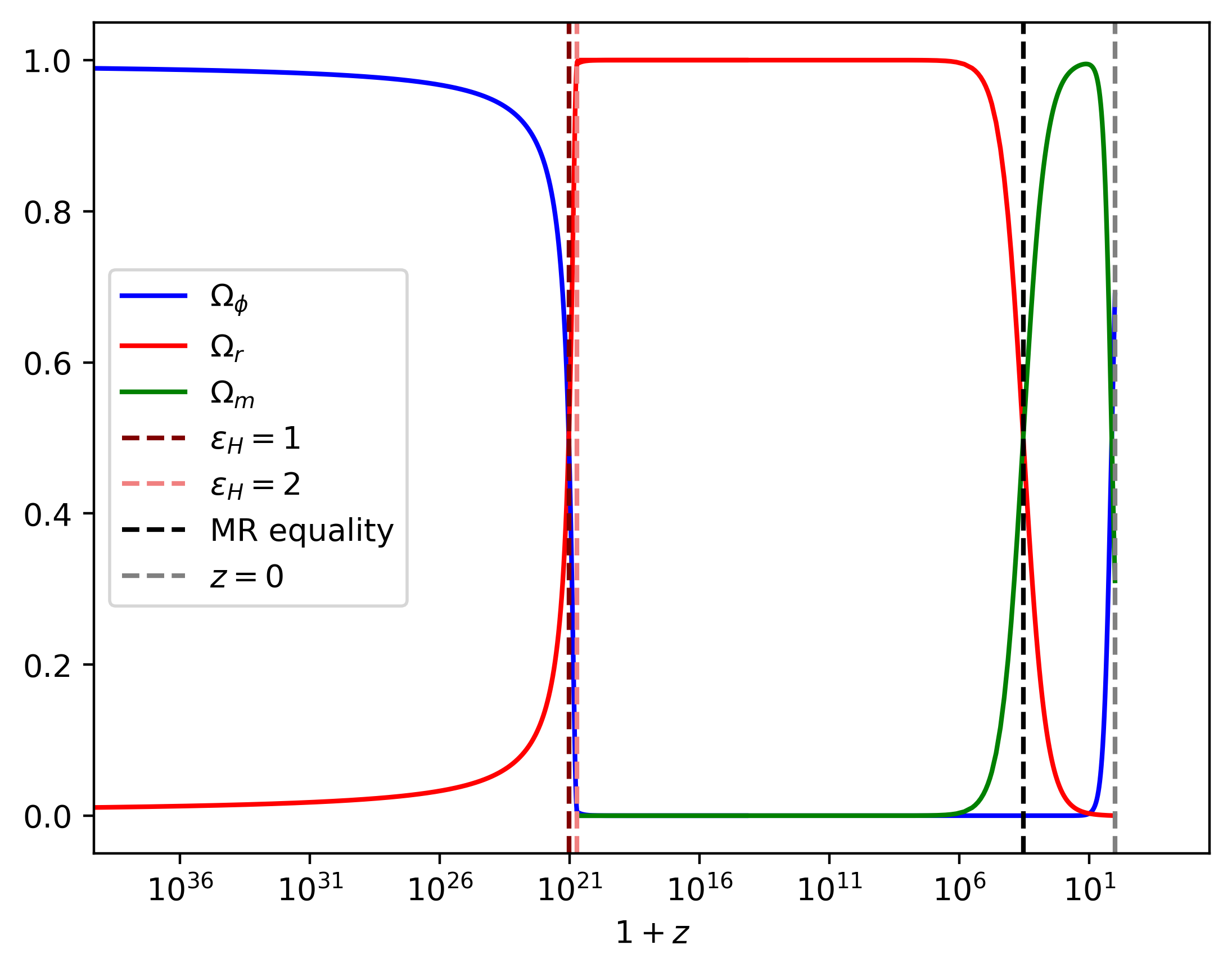}
    \caption{Evolution of the density parameters $\Omega_\phi$, $\Omega_r$ and $\Omega_m$}
    \label{density-params}
\end{figure}

\begin{figure}[ht!]
    \centering
    \includegraphics[width=0.48\textwidth]{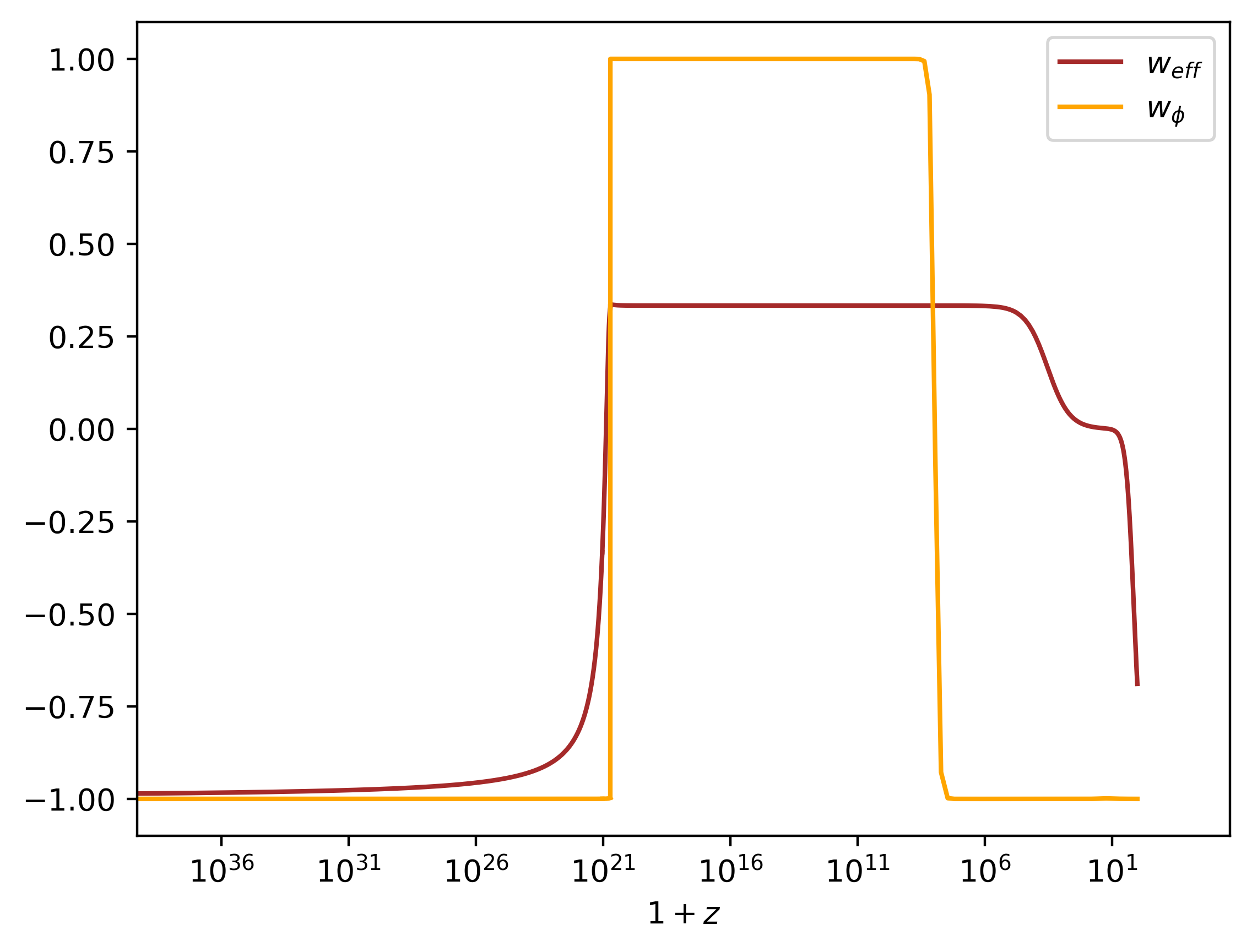}
    \caption{Evolution of the equation of states $\omega_\phi$ and $\omega_{\rm eff}$}
    \label{eos-evol}
\end{figure}

With these arrangements, the model successfully displays thawing Dark Energy at late times. The evolutions of all the density parameters ($\Omega_\phi$, $\Omega_r$ and $\Omega_m$) throughout the evolution are depicted in Fig.~\ref{density-params} and the evolutions of the equation of state of the quintessential inflaton field and the effective equation of state are shown in Fig.~\ref{eos-evol}. We have chosen $V_1=3.64\times10^{-119}\,m_p^4$ and $C_m=1.37\times 10^{-87}\,m_p^2$ to obtain the current value of vacuum energy and the slow-roll of the quintessence field respectively. The evolution yields $z_{\rm eq}=3388$ and $\Omega_{\phi_0}=0.689$ and $\Omega_{m_0}=0.311$. Moreover, the choice of the value of $C_m$ yields $\Upsilon_m/\Upsilon_r\sim 10^{-56}$ at the beginning of the radiation epoch ensuring the ineffectiveness of $\Upsilon_m$ during inflation as was anticipated before.

We would like to highlight that in this setup, once the quintessence field rolls back to the bottom of the Witten-O'Raifeartaigh potential from the right wing, the accelerating expansion will cease. Hence, our model depicts a transient Dark Energy epoch. The duration of this Dark Energy epoch will clearly depend upon how slowly the quintessence field rolls back to the bottom of the potential, or in other words, how strong the the dissipative coefficient $\Upsilon_m$ is. Larger the dissipation longer will be the duration of the Dark Energy epoch. Thawing quintessence models, as the one studied here, are in good agreement \cite{deSouza:2025rhv} with the recent DESI results \cite{DESI:2025zgx}, and, moreover, a transient Dark Energy epoch doesn't suffer from the trans-Planckian issues that concerns the ever-lasting Dark Energy era \cite{deSouza:2025rhv}. Hence, the quintessence model presented here is observationally sound as well as devoid of issues like trans-Plankian problem.

\section{Discussion and Conclusion}
\label{conclusion}

The Witten-O'Raifeartaigh potential \cite{Witten:1981kv, ORaifeartaigh:1975nky, Martin:2013tda}, having two distinct wings, a very steep left wing and a flatter right wing, opens up avenues that can be exploited in  the cosmological context. This potential is also theoretically well-motivated, arising within the framework of supersymmetry and supergravity, which have prospects for UV completion. Though, previously it was shown that CI can successfully take place in the flatter right wing of the potential \cite{Martin:2013tda}, no cosmologically relevant dynamics could be realized in the context of either inflation or dark energy in the left wing of the potential, since it is too steep to sustain slow-roll dynamics. 

In this paper, we  demonstrated that WI in the strong dissipative regime can take place in the steep left wing of the Witten-O'Raifeartaigh potential as slow roll can be maintained despite the steepness of the potential by the means of the additional friction arising due to persistent strong dissipation of the inflaton field. We  also verified the observational validity of this model through a thorough MCMC analysis with the recent datasets (by a joint analysis of Planck, ACT and DESI \cite{ACT:2025fju, ACT:2025tim}). This is the main highlight of this paper. Furthermore, once we have established that the steeper left wing can yield WI successfully, the flatter right wing becomes  potentially available for another slow-roll phase that can play the role of quintessence Dark Energy sourcing the late-time accelerated expansion of the Universe.

However,  we realized that the desired quintessence dynamics in the flatter right wing cannot be realized naturally unless a few additional accommodation is accounted for. First of all, it is essential to consider two different normalizations for the two wings of the Witten-O'Raifeartaigh potential in order to bridge  the  extreme hierarchy of energy scales between inflation and  late-time  acceleration.  The requirement of  two different normalizations spoils the smoothness of the Witten-O'Raifeartaigh potential, making it a $C^1$ curve, though this doesn't hinder the quintessential inflation dynamics to take place. Secondly,  persistent strong dissipation of the inflaton field during WI drains away a significant fraction of the kinetic energy of the inflaton,  so much so that it fails to climb up far enough along  the right wing to a region where the potential is sufficiently flat for sustaining a slow-roll quintessence phase. In fact, the scalar field only manages to climb up to a small distance in the field space where the right wing is still too steep to sustain slow roll. In order to alleviate this difficulty, a WI-like dissipative dynamics of the quintessence field has to be considered in the right wing as well so that slow-roll quintessence dynamics can be ascertained  in the right. Nevertheless, such a setup in which the scalar field dissipates during both the accelerated epochs, establishes a similarity between the dynamics of the quintessential inflaton field in these two epochs of acceleration. The dissipation of the scalar field is well justified as long as the scalar field has substantial couplings with other degrees of freedom, {which is} natural to consider in any particle physics model. 

With these accommodations, we demonstrated that the `broken' Witten-O'Raifeartaigh potential can indeed facilitate a successful quintessential inflationary dynamics where dissipation of the scalar field, a defining feature of WI, plays a major role  during both the accelerated epochs. We also showed,  in this whole setup, that  (i) the warm inflationary regime is well in accordance with observations, (ii) the setup doesn't  require any conventional or variant reheating epoch, (iii) the setup yields a transient  (thawing) Dark Energy epoch.

\acknowledgements
The work of S.D. is supported by the Start-up Research Grant (SRG) awarded by Anusandhan National Research
Foundation (ANRF), Department of Science and Technology, Government of India [File No. SRG/2023/000101/PMS]. 
UK warmly acknowledges the Axis Bank PhD program at Ashoka University for PhD fellowships provided by Axis Bank. UK also acknowledges the HPC facility at Ashoka University (Chanakya@Ashoka) for providing computational resources essential to this work. Additionally, UK is grateful to Kandaswamy Subramanian for insightful discussions that have contributed significantly to the development of this work. SSM is supported by IBS under the project code, IBS-R018-D3. SSM was also supported by the STFC Consolidated Grant [ST/T000732/1] at the University of Nottingham. SSM  gratefully acknowledges support from the IUCAA Visitor Academic Programme. VS thanks the Anusandhan National Research Foundation (ANRF), India,
for the National Science Chair Professorship which provided partial funding for this work.
SD would like to thank Rudnei Ramos for many useful discussions over the years which have been very useful in this work.


\label{Bibliography}
\bibliography{refs}


\end{document}